# The multifractal load balancing method

Ivanisenko Igor
Dept. of Applied Mathematics
Kharkiv National University of Radio Electronics
Kharkiv, Ukraine
ivanisenko79@yahoo.com

Radivilova Tamara
Dept. of Telecommunication Systems
Kharkiv National University of Radio Electronics
Kharkiv, Ukraine
tamara.radivilova@gmail.com

*Abstract*—The load-balancing system, built on the basis of a subsystem load balancer and subsystem control and monitoring that closely interact with each other was propose in work. This system is presented as a queuing system with priority service discipline. In the described queuing system parallel processing flow applications is occurs in the multiple serving devices and successive junction them into unified stream. The method of multifractal load balancing is submited on the basis of the developed system of load balancing.

## I. Introduction

The specificity of the approach to information processing and storage, based on distributed systems, is the fact that the overall availability of resources to different user tasks. Modern high balancers are heterogeneous multifunctional devices oriented to processing various types of data [1,2]. The use of different types matchers allows to serve a large range of different data streams.

In this work the functioning of the load balancer, which has a self-similar properties is considered. The reason of self-similar properties of traffic is a variety of data, packets, features of the distribution of files on the servers, their size, the typical user behavior. Self-similar load causes major delays and packet loss, even if the total intensity of all flows is distant the maximum permissible value [2-4].

In recent years, the properties of multifractal traffic is investigated. Multifractal traffic is defined as the expansion of self-similar traffic due to considering of scalable properties of the statistical higher orders characteristics. Methods of management multifractal traffic techniques began actively explored to improve services network, in particular, the selection and application of load balancing methods and algorithms [1-3,5-7].

However, though the growing number of studies in this direction, a series of questions is still open. These include studies the improving of quality of services mechanisms and load balancing methods in distributed systems [1,5-7].

## II. Load balansing system

Multifractal load balancing system (figure 1) is proposed to build on the basis of subsystems that cooperate with each other closely [1,6,7]:

- load balancer subsystem: load balancing algorithm, information about the current state of the system, flexible configuration QoS, dynamic allocation of traffic over various communication channels and nodes based on their current state, loading extend, the administrative load-balancing policies.

- control and monitoring subsystem: to collect and analyze statistics about current state of the system, finding the multifractal properties of the incoming data stream, the calculation of flow distribution in the network nodes based on traffic classification and utilization of servers and communication channels.

Servers send information about their load to balancer.

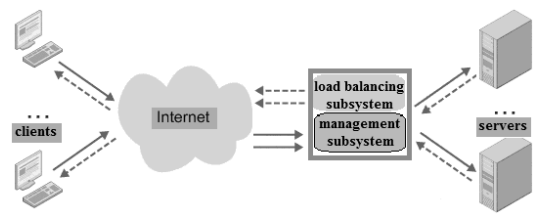

Fig. 1. The scheme of distributed network

The system executes the following scenario works: dynamic balancing of the multifractal traffic based on QoS parameters through the communication channels between servers based on their current workload and the actual bandwidth at any given time.

It is possible to use special queuing system (QS) for modern high-speed simulation of distributed systems, where data processing performed on several independent computing devices [1,6]. The parallel processing of multifractal applications flow by multiple devices and consequent merging them into a single stream will allow to describe such a system. Thereby, traffic management in the load balancer can be described by using QS.

## III. Load balancer as QS

The load balancer is described as a multi-channel QS with limited storage capacity, shown in Figure 2. The load balancing subsystem is a block recource, and control and monitoring subsystem is a block service procedure. The input QS received several independent multifractal applications flows (packages) with varying intensity $\lambda 1, \lambda 2, \ldots, \lambda N$. Queuing time is dependent on the type of stream and a type of the service node. Priority service discipline (PSD) queue requirements based on the account of the importance of service requirements [1,6]. While all priority service requests will not be processed, packages of other types stay in queue until the end of their lifetime. Newly received priority requests dropping off the process non-priority ones and with a probability equal to one displace them in storage (if have free waiting space), or outside the system (if the storage is full). Packages displaced from service are join to queue of non-priority requirements and can be serviced after all the priority ones. The drives are separate, free space fully accessible for all newly received requests. Unlike typical priority QS the considered system is equipped with probabilistic eject mechanism. Priority package, that found all places busy during of processing other priority packet, is replacing one of the lower priority packets from the drive with a given probability, and takes his place. Displaced package is lost or sent back to the queue.

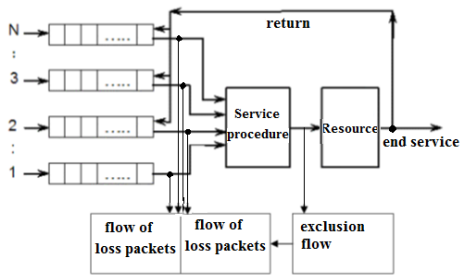

Fig. 2. The scheme of priority QS

Queues and losses, that generated by traffic with multifractal properties, depend on the characteristics of the multifractal: generalized Hurst exponent function. Value range of the generalized Hurst exponent corresponds to the power of heterogeneity of traffic, i.e. it characterizes the scatter of the data [2,3]. The value of the ordinary Hurst parameter, obtained from generalized index, corresponds to power of long-term dependence of realization and characterizes the correlation properties of traffic.

## IV. THE METHOD OF LOAD BALANCING

The paper proposes the following dynamic dependent on content the method of load balancing with a modified feedback:

*1) in the traffic that input to the switch, select the window X of the fixed length T;*

*2) find the selective value of the function of the generalized the Hurst exponent h(q), the value of Hurst exponent H=h(2) and the range of values of the generalized Hurst exponent $\Delta h = h(q_{min}) - h(q_{max})$ for the area of traffic in the dedicated window;*

*3) collect and analyze statistical data: the intensity of the incoming flows λ1,λ2,...,λN, state of servers Uj={Upj,Umj,Udj} (CPU utilization Upj, memory Umj, disk Udj, j-th server, created by i-th class of requests);*

*4) on the basis of multifractal properties of traffic (values of p. 2), and the traffic intensity calculate necessary amount of resources for each i-th traffic classes;*

*5) perform calculations of flow distribution of network nodes based on traffic classification and utilization of servers and communication channels. On the basis of the data obtained is forecasting workload of the servers in the next step;*

*6) balancing traffic across the servers, according to the load balancing algorithm considering each class of flows;*

*7) perform balancing of underestimating the forecasting number of requests. Reassessment of is not considered an algorithm because it does not introduce any significant;*

*8) collect data about servers utilization Uj and send it to the system load balancing to calculate the new flow distribution;*

*9) move the the window X of the fixed length T forward by a predetermined shift amount ΔT;*

*10) carry out the traffic analysis and forecast of the next value of server load.*

The proposed load balancing method turned to providing a statistically uniform distribution of the load on servers, high performance, capacity, fault tolerance (automatically detecting failures of nodes and redirecting the flow of data among the remaining) and low response time, the quantity of service information and losses.

## CONCLUSION

Described in this paper load balancing system, built on the basis of subsystem load balancer and subsystem control and monitoring that closely interact with each other. The system of load balancing is presented as a QS with priority service discipline. Parallel processing priority flows applications occur in multiple serving devices and then packing them into unite stream that described in the QS. On the basis of the developed load balancing system is provided a method of multifractal load balancing. In the further work is necessary to assess the effectiveness of the developed method of load balancing based on multifractal properties of the traffic in the distributed network possessing the various dynamic load balancing algorithms. Evaluate the effectiveness of the proposed method is expedient by the following parameters: bandwidth, cost, response time and resource utilization.